\documentclass{JHEP3}

\usepackage[latin1]{inputenc}
\usepackage{amsmath,amssymb,amsfonts,epsfig} 
\usepackage{bbm,bm}
\usepackage{graphicx}

\newcommand{\Eqref}[1]{Eq.~\eqref{#1}}

\title{Axion-like-particle search with high-intensity lasers}

\author{Babette D\"obrich and Holger Gies\\
  Theoretisch-Physikalisches Institut, Friedrich-Schiller-Universit{\"a}t
Jena,\\
\& Helmholtz Institute Jena, Max-Wien-Platz 1, D-07743 Jena, Germany\\
   E-mail: \email{babette.doebrich@uni-jena.de, holger.gies@uni-jena.de}}
\received{}             
\revised{}
\accepted{}             

\preprint{}             

\abstract{We study ALP-photon-conversion within strong inhomogeneous electromagnetic fields as provided by contemporary high-intensity laser systems.
We observe that probe photons traversing the focal spot of a superposition of Gaussian beams of a single high-intensity laser at fundamental
and frequency-doubled mode can experience a frequency shift due to their intermittent propagation as axion-like-particles.
This process is strongly peaked for resonant masses on the order of the involved laser frequencies.
Purely laser-based experiments in optical setups are sensitive to ALPs in the $\mathrm{eV}$ mass range and can thus complement ALP searches at dipole magnets.
}


\begin{document}


\section{Introduction}
{
(Pseudo-)scalar particles or corresponding bound states are often intimately
related with the realization of global symmetries in particle physics. They
can play the role of condensation channels of symmetry-breaking condensates as
well as occur as (pseudo-)Goldstone bosons. A particularly prominent example
is given by the axion, the pseudo-scalar pseudo-Goldstone boson of a
conjectured Peccei-Quinn symmetry \cite{Peccei:1977hh}, which so far provides
for the only viable solution to the strong CP problem. 

Even though the axion or more generally, axion-like-particles (ALPs),
generically develop couplings to photons by means of a dimension-5 operator,
giving rise to an effective action of the type
\begin{equation}
\mathcal{L}_{\mathrm{P}}=-\frac{1}{4}F_{\mu\nu}F^{\mu\nu}+\frac{1}{2}\partial_{\mu}\phi\partial^{\mu}\phi-\frac{1}{2}m^{2}\phi^2+\frac{1}{4}g\phi F_{\mu\nu}\tilde{F}^{\mu\nu}
\ , \label{eq:general_lag}
\end{equation}
where, $\phi$ is the axion-like field with mass $m$ and coupling $g$, direct
axion searches are inflicted by a presumably extremely small coupling
$g$. Strongest bounds on this coupling for a wide mass range exist typically in the
sub-eV regime and are provided by astrophysical arguments related to stellar
cooling or direct solar observation \cite{CAST}, constraining $g$ to lie below
$g\lesssim 10^{-10}\text{GeV}^{-1}$. 

As the relevant stellar axion generation processes involve momentum transfers
in the keV range, these bounds are somewhat  model dependent and do not
immediately apply to general (pseudo-)scalar particles with couplings (and
corresponding form factors) becoming sizable only at much lower momentum
transfers \cite{Jaeckel:2006xm}. This has inspired the realization of many
laboratory experiments based on optical probing which are specifically
sensitive to couplings to  ALPs at low momentum transfer, typically at
$|q|\simeq \mathcal{O}(1\mu\text{eV})$.  Most commonly,
laboratory searches for ALPs are based on the mixing 
between axions and photons induced by a macroscopic external magnetic field
through the trilinear coupling 
in Eq. \eqref{eq:general_lag}. 
For example, in polarimetric measurements \cite{Maiani:1986md}, one can determine the ellipticity and rotation of a polarized light beam induced within an external field and thereby explore
the parameter space of the ALP mass and coupling. This mechanism has been exploited e.g. at the ALP search at PVLAS \cite{Zavattini:2007ee} and is also planned at BMV \cite{battesti}.
Further, photon production in strongly inhomogeneous fields \cite{Guendelman:2009kv} as well as high-sensitivity interferometric measurements \cite{Dobrich:2009kd} have recently
been suggested as means of ALP detection.

Alternatively, one can utilize the weak coupling between ALPs and ordinary matter in order to shine photons in external fields through light-blocking walls,
in so-called ''Light-shining-through-walls'' (LSW) setups \cite{Sikivie:1983ip}. This is possible if probe photons are converted into ALPs in front of the light-blocking wall
and reconverted into photons behind that wall, where in typical laboratory searches the conversion processes are induced by strong dipole magnets.
Such searches are currently performed e.g. by the ALPS \cite{Ehret:2010mh},
LIPSS \cite{Afanasev:2006cv}, GammeV \cite{Steffen:2009sc}, BMV \cite{Robilliard:2007bq} and OSQAR
\cite{OSQAR} collaborations, also aiming at cosmologically relevant scalar
fields \cite{Gies:2007su}  or hidden gauge bosons \cite{Ahlers:2007qf}.

In both cases, polarimetry as well as LSW, the decisive parameter for the best obtainable bounds on the ALP mass and coupling is the product of the field strength of the
external magnetic field $B$ and its spatial extent $L$ being a measure for the optical path length.
Typically, the dipole magnets which are employed in these setups provide field strengths of $B\sim\mathcal{O}(1-10)\mathrm{T}$ extending over a length of $L\sim\mathcal{O}(1-10)\mathrm{m}$.
By use of cavities for the probe beam, the interaction region can be extended by a few orders of magnitude, depending on the details of the setup.
Also, the idea of resonant enhancement in LSW setups  \cite{Sikivie:2007qm} holds out the prospect of considerably enlarging the effective interaction region.

On the other hand, the highest field strengths which are obtainable nowadays in a laboratory are present within the focal spots of high-intensity laser systems.
Current Multi-Terawatt lasers achieve peak field strengths of $B\sim\mathcal{O}(10^5-10^6)\mathrm{T}$,
however, naturally at cost of the spatial extent of these fields, which ranges from $L\sim\mathcal{O}(1-10)\mu\mathrm{m}$. 
Nevertheless, the parameter $BL$ in the laser focus lies within the same ball park as for the dipole searches,
calling for proposals of ALP search within high-intensity laser-based setups
\cite{Gies:2008wv}. 

In addition, the achievable laser intensity has gone up by more than six orders of magnitude since the invention of chirped pulse amplification \cite{CPA},
with the prospect that the parameter $BL$ within planned facilities such as ELI \cite{ELI} will considerably exceed the equivalent parameter at dipoles within the near future.
This makes high-intensity lasers
a dedicated tool \cite{Heinzl:2008an} for fundamentals test of QED nonlinearities \cite{Heisenberg:1936qt} most prominently comprising e.g. the possible detection of 
vacuum birefringence \cite{Heinzl:2006xc,Tommasini:2010fb,DiPiazza:2006pr,King:2009,Homma:2010jc} and the Schwinger effect \cite{Dunne:2008kc}.
Strikingly, these searches implicitly constitute also a probe for physics
beyond the Standard model of particle physics
\cite{Gies:2008wv,Jaeckel:2010ni,Tommasini:2009nh}.

At first sight, however, the above mentioned optical techniques, namely polarimetry and LSW setups for ALP detection seem to be obstructed by the nature of the electromagnetic field
configuration within high-intensity lasers. High intensities and thus high field strengths can only be attained by pulsed lasers with typical pulse lengths of
$\tau\sim\mathcal{O}(10-100)\mathrm{fs}$ and repetition rates of $f_{\mathrm{rep}}\lesssim1\mathrm{Hz}$.  Thus, the cavity enhancements which are used for
polarimetric measurements at dipoles are not easily available for purely laser-based setups.

On the other hand, also the insertion of  a light-blocking wall in a purely laser-based setup is disfavored: In order to avoid damaging of the wall by the high-intensity lasers,
the two focal spots for the conversion and reconversion processes would in practice be required to be separated by $\mathcal{O}(\mathrm{cm})$.
As the spatial extent of the focal spots of lasers is by orders of magnitude smaller, the generic angular spread of the beam of ALPs released from the first focal spot would
significantly reduce the number of ALPs that could possibly hit the second spot for reconversion.
Even if the angular spread could be minimized, a laser-based LSW experiment would demand for a temporally very well-synchronized setup.

In this work, we suggest another mechanism for high-intensity lasers which does neither rely on polarimetry nor on light-blocking walls.
Since the electromagnetic field provided by the high-intensity lasers varies at a scale which can be of the same order of magnitude as the wave length of the probe photon,
this photon can experience a frequency shift when traversing the focal region of the external field  owing to the nonlinear term in Eq. \eqref{eq:general_lag}.
Consequently, the detection of such frequency shifted photons could thus point towards the existence of ALPs.
In the following, we compute this effect quantitatively and discuss the required setup and specifications of lasers which are necessary for its detection.

The paper is organized as follows: In Sects.~\ref{sec:EOM} and
\ref{sec:parameterization}, we first give the equations of motion for the ALPs
and probe photons which we reduce to one spatial dimension for simplicity
where the formalism developed in \cite{Adler:2008gk} can be applied and
discuss the necessary parameterization of the high-intensity laser beams.  In
Sects.~\ref{sec:pho_ax_con} and \ref{sec:ax_pho_con} , we compute the
photon-axion conversion and back-conversion amplitudes in a specific laser
configuration and discuss the physical reasons why this setup can lead to a
frequency shift for the probe photons.  Finally, in Sect.~\ref{sec:results},
we will summarize our findings and estimate exclusion bounds achievable for
the operational high-intensity laser facility at the Institute of Optics and
Quantum Electronics in Jena \cite{IOQ} and the planned Exawatt facility ELI
\cite{ELI}.  }

\section{Equations of motion \label{sec:EOM}}

As we are interested in the effects of the nonlinear interaction of laser
photons, let us give the equations of motion for the photon and the axion
field that follow from Eq.\eqref{eq:general_lag}:
\begin{eqnarray}
\partial_{\mu}\partial^{\mu}\phi+m^2\phi-\frac{1}{4}g F_{\mu\nu}\tilde{F}^{\mu\nu} &=& 0 \label{eq:axion_EOM} \\
\partial_{\mu}F^{\mu\nu}-g(\partial_{\mu}\phi)\tilde{F}^{\mu\nu} &=& 0 \label{eq:photon_EOM}\ .
\end{eqnarray}
Below, we study these equations in a rather general setup, assuming the
interaction of three independent electromagnetic fields, which we all presume
to be provided by high-intensity lasers. As it will turn out later,
experimentally there is actually just the need for at most two sources.

We split up the field strength tensors into contributions of a probe beam $a^{\mu}_{\mathrm{in}}$ and two external fields $A^{\mu}_j$ and $A^{\mu}_k$;
in addition, we neglect self-interactions of these fields.
By coupling to $A^{\mu}_j$, the probe photons can be converted into ALPs, see Eq. \eqref{eq:axion_EOM}.
Successively, in Eq. \eqref{eq:photon_EOM}, these axions can be reconverted into photons $a^{\mu}_{\mathrm{out}}$ through a field  $A^{\mu}_k$.
For simplicity, we specialize to a one-dimensional setup.
In addition, since we do not intend to focus on polarimetry later on, we assume the incoming probe photons to be polarized along the $y$ axis and
to propagate along the positive $z$ axis without loss of generality.

Under these presumptions, employing Coulomb gauge and the metric $g=(+,-,-,-)$, Eqs. \eqref{eq:axion_EOM} and \eqref{eq:photon_EOM} can now be written as

\begin{eqnarray}
(\partial_t^2-\partial_z^2+m^2)\phi(z,t) &=& -g\left[e_{\mathrm{in}}^y(z,t) B_j^y(z,t) + b_{\mathrm{in}}^x(z,t) E_j^x (z,t)\right] \label{eq:axion_eom_1d}  \\
(\partial_t^2-\partial_z^2)a_{\mathrm{out}}^y(z,t) &=& -g\left[B_k^y(z,t)\partial_t \phi(z,t) + E_k^x(z,t) \partial_z \phi(z,t) \right] \label{eq:photon_eom_1d} \ .
\end{eqnarray}
For the external fields, we consider different cases of propagation along the $\pm z$ axis or orthogonal to the $z$ axis.
In the first case of parallel propagation the external lasers interact through both their electric and magnetic field components.
In the second case of orthogonal propagation, the external fields can couple only through either their electric or magnetic field component. 

In particular we see from Eq. \eqref{eq:axion_eom_1d} that the axion amplitude vanishes trivially in a setup where the fields $e_{\mathrm{in}}^y$ and $E_j^x$ propagate both along the $+z$ axis,
since then $e_{\mathrm{in}}^y=-b_{\mathrm{in}}^x$ and $E_j^x=B_j^y$.
By contrast, for counter-propagation of the two beams, the fields on the right-hand side of Eq. \eqref{eq:axion_eom_1d} add up, since then we have $E_j^x=-B_j^y$.

If the propagation axis of the external field lies orthogonal to the $z$ axis, either the magnetic or the electric field component can mediate the conversion process.
However, the axion amplitude in Eq. \eqref{eq:axion_eom_1d} is invariant under this choice for a linearly polarized external beam.
On the other hand, the back-conversion from the axions into photons is not independent of whether the axion couples to the electric or magnetic field component,
due to the asymmetric coupling structure in Eq. \eqref{eq:photon_eom_1d}. This is due to the pseudoscalar nature of $\phi$. If we had taken $\phi$ to be a scalar field,
then the coupling structure in Eq. \eqref{eq:photon_eom_1d} would be given by interchanging $B_k^y$ and $ E_k^x$.

In the following, we solve Eqs. \eqref{eq:axion_eom_1d} and
\eqref{eq:photon_eom_1d} using the retarded Green's functions along the lines
of \cite{Adler:2008gk} for the massive and massless differential operators in
the equations of motion,
\begin{eqnarray}
G_{m}^{\mathrm{R}}(z,t)&=& \frac{1}{2} J_{0}\left(m\sqrt{t^2-|z|^2}\right)\theta\left(t-|z|\right) \label{eq:Gm} \\
G_{0}^{\mathrm{R}}(z,t)&=& \frac{1}{2}\theta\left(t-|z|\right) \label{eq:G0} \ ,
\end{eqnarray}
respectively.

The solutions to Eqs. \eqref{eq:axion_eom_1d}  and \eqref{eq:photon_eom_1d} naturally depend on the details of the kinematic setup.
In the following, we choose with hindsight a specific setting for which the conversion process of photons into axions and vice versa leads
ultimately to a frequency shift of the probe beam. If this frequency shift exceeds the natural line width of the probe beam, it can constitute a measurable signal,
possibly indicating the existence of ALPs.

\section{Parameterization of the fields \label{sec:parameterization}}

In order to solve Eqs. \eqref{eq:axion_eom_1d}  and \eqref{eq:photon_eom_1d} we need a parameterization for the electric and magnetic fields of the three laser beams.
A good model for the spatial inhomogeneities of a focused beam is given by Gaussian beams \cite{Davis:1979}, which are solutions to the paraxial wave equation.

In consistency with our one-dimensional model, we restrict ourselves to the lowest-order contribution in the aspect ratio $\theta_0=\frac{w_0}{z_r}$, where $w_0$ is the waist size
and $z_r$ the Rayleigh length of the beam\footnote{At higher orders in the aspect ratio, the beam acquires polarization components which are neglected in the one-dimensional calculation
and loses transversality.}. The waist size of a Gaussian beam is a measure for the transversal extent of the beam at the focus, whereas the Rayleigh
length parameterizes the broadening of the focus along the propagation direction, cf. Fig. \ref{fig:setup}.
They are related through the wavelength of the beam as $z_r=\frac{\pi w_0^2}{\lambda}$.
To maximize the interaction of the lasers, their focal spots should have a sizable overlap. Here, we assume the waist size to be minimized at the origin $x=y=z=0$ for the probe beam
as well as for the two external fields.

Note that for Gaussian beams, the existence of the peak external field strength is, of course, also limited by a temporal pulse length $\tau$.
This scale must be larger than the time it takes the probe photons to traverse the external fields\footnote{Here and in the following $c=\hbar=1$.}:
$\tau_{\mathrm{ext}}\gtrsim z_r^{\mathrm{ext}}$ and $\tau_{\mathrm{ext}}\gtrsim w_0^{\mathrm{ext}}$.
In the present study, we formally work in the limit of infinite pulse length for both external field and probe beam: $\tau_{\mathrm{ext}}$, $\tau_{\mathrm{in}}\rightarrow\infty$.
Our final result will thus be phrased in terms of a transition probability for the photons of the probe field. In practice, as real facilities are limited in energy, intensity and power,
an optimization of the effect under such constraints can typically be expected for all pulse parameters of probe and external field being roughly of the same order.

\FIGURE{
\centerline{
\includegraphics[scale=0.45]{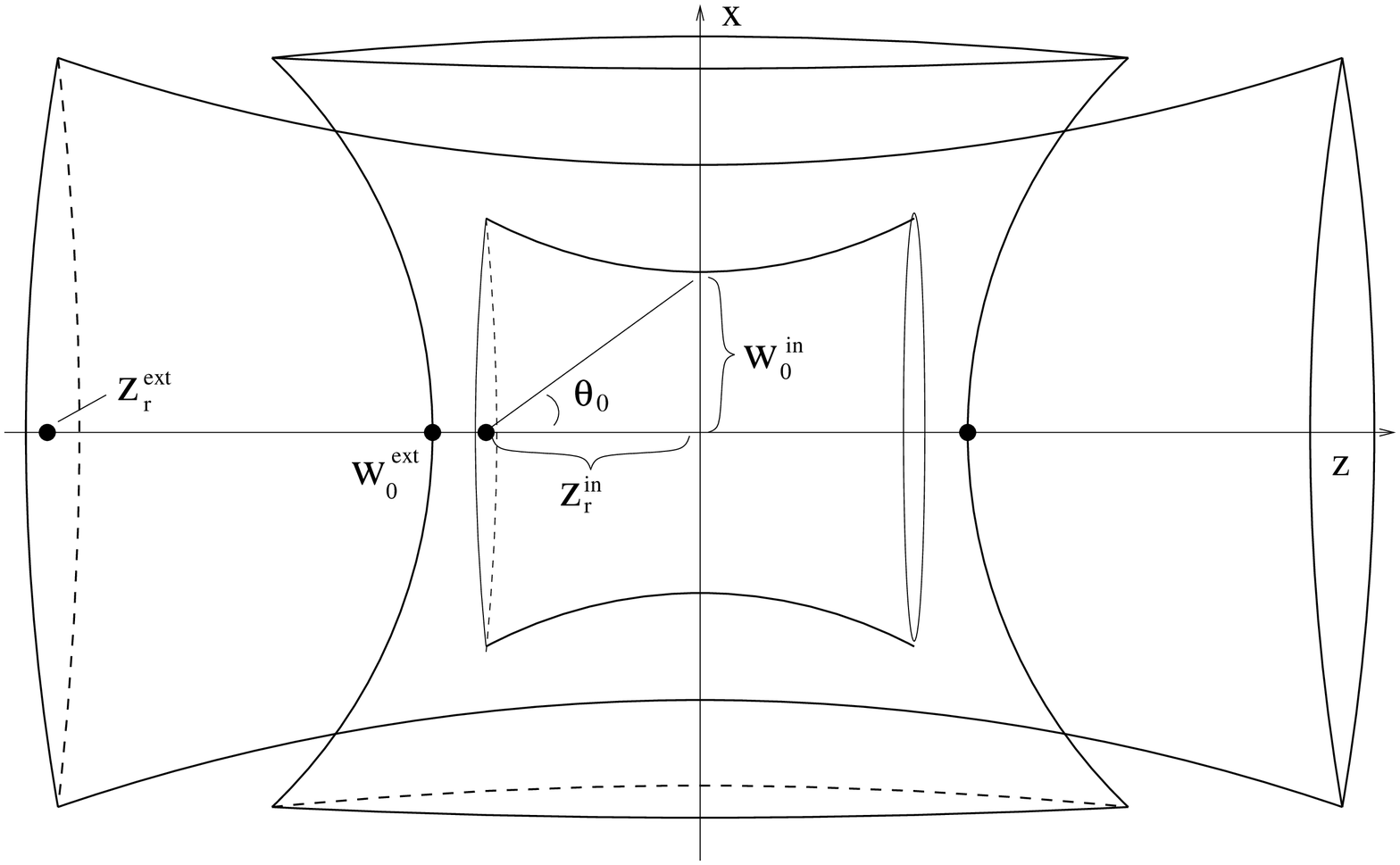}}

\caption{Spatial overlap of three focused laser beams at the coordinate center $x=y=z=0$. In this picture, the innermost beam gives the Gaussian probe beam
embedded in two external beams which propagate orthogonally and transversally to it, respectively. The waist size $w_0^{\mathrm{ext}}$ as well as the Rayleigh lengths
$z_r^{\mathrm{in}}$ and  $z_r^{\mathrm{ext}}$ constitute scales for the extent of the beam foci along the propagation direction of the probe beam and thus  parameterize the
drop-off of the electric and magnetic field components of the lasers. The parameter $\theta_0=\frac{w_0}{z_r}$ is the aspect ratio for which typically $\theta_0\ll 1$.}

\label{fig:setup}
}

Under these presumptions, the electric and magnetic field for the incoming probe beam propagating along the positive $z$ axis reads
 
\begin{equation}
e_{\mathrm{in}}^y(z,t)=-b_{\mathrm{in}}^x(z,t)=\frac{E_{\mathrm{in}} }{\sqrt{1+(z/z_r^{\mathrm{in}})^2}}\sin
\left(\omega_{\mathrm{in}} t- k_{\mathrm{in}} z+\arctan\left(\frac{z}{z_r^{\mathrm{in}}}\right)\right)\ , \label{eq:probe_model}
\end{equation}
where $E_{\mathrm{in}}$ is the amplitude of the field, $\omega_{\mathrm{in}}$ the frequency, $k_{\mathrm{in}}$ the wave vector and $z_r^{\mathrm{in}}$ the Rayleigh
length of the probe beam, as discussed above. From Eq. \eqref{eq:probe_model} it can be seen that the Rayleigh length not only characterizes the longitudinal extent of the field,
but also appears in the so-called Gouy phase shift that a focused light beam undergoes, when passing through its focus \cite{Gouy}.
In addition, it holds in Eq. \eqref{eq:probe_model} that $\omega_{\mathrm{in}}=k_{\mathrm{in}}$ in vacuum; nevertheless, we stick to this notational distinction, since it will
simplify the discussion of energy and momentum conservation later on. Lastly, the probe beam model in Eq. \eqref{eq:probe_model} as well as the external beams can in general 
include also a constant phase factor. However, since only the relative phase between the three beams is important, we omit such a phase factor in the above definition.

\section{Photon-Axion-Conversion \label{sec:pho_ax_con}}

We can now calculate the photon-axion conversion by solving Eq. \eqref{eq:axion_eom_1d}.
With hindsight, we choose the external field to propagate orthogonal ($\bot$) to the $z$ axis and discuss the implications of other possible settings later.

Without loss of generality, we choose the external field to couple through its electric field component, and plug in its Gaussian beam form
\begin{eqnarray}
E_{j}^{x}(z,t)&=& E_{\bot}\sin(\omega_{\bot} t+\psi_{\bot}) e^{-(z/w_0^{\bot})^2}, \label{eq:E_j_field} \\
B_{j}^{y}(z,t)&=& 0\ , \label{eq:B_j_field}
\end{eqnarray}
where $ E_{\bot}$ is the amplitude, $\omega_{\bot}$ the frequency, $w_0^{\bot}$ the waist size and $\psi_{\bot}$ the phase of the external beam.
Thus, combining the ALP equation of motion, \eqref{eq:axion_eom_1d}  with the Green's function of Eq. \eqref{eq:Gm} and the parameterization of the fields,
Eqs. \eqref{eq:probe_model} and \eqref{eq:E_j_field}, we obtain for the axion field:
\begin{multline}
\phi(z',t')=\frac{1}{2}g E_{\mathrm{in}} E_{\bot} \int_{-\infty}^{\infty} \mathrm{d}z'' \frac{1}{\sqrt{1+(z''/z_r^{\mathrm{in}})^2}} e^{-(z''/w_0^{\bot})^2} \\
\times \int_{-\infty}^{\infty} 
\mathrm{d}t'' J_{0}\left(m\sqrt{(t'-t'')^2-|z'-z''|^2}\right) \theta\left((t'-t'')-|z'-z''|\right) \\
\times \sin(\omega_{\mathrm{in}} t''- k_{\mathrm{in}}z''+
\arctan(\frac{z''}{z_r^{\mathrm{in}}}))\sin(\omega_{\bot} t'' +\psi_{\bot})\ . \label{eq:phi_int_int}
\end{multline}
Here we have used primed arguments for the ALP field $\phi$, in order to indicate that these are variables over which we still have to integrate in the back-conversion process later on.

As typical laboratory scales are many orders of magnitude larger than the spatial extents of the beams, it is justified to perform the $z''$
integration from $-\infty$ to $\infty$. Following the constraints for the respective pulse lengths $\tau_{\mathrm{ext}}$ and $\tau_{\mathrm{in}}$, which were discussed in the previous section,
we let also the integration over $t''$ extend from $-\infty$ to $\infty$ for computational simplicity.

We rewrite the sines as sum of exponentials and substitute $t''\rightarrow t'-T$ in Eq. \eqref{eq:phi_int_int}, yielding
\begin{multline}
\phi(z',t')=-\frac{1}{8}g E_{\mathrm{in}} E_{\bot} \int_{-\infty}^{\infty} \mathrm{d}z'' \frac{1}{\sqrt{1+(z''/z_r^{\mathrm{in}})^2}}e^{-(z''/w_0^{\bot})^2}\\
\Biggl[e^{-i k_{\mathrm{in}}z''}e^{i(\omega_{\mathrm{in}}+\omega_{\bot})t'} e^{i(\arctan(z''/z_r^{\mathrm{in}})+\psi_{\bot})} \int_{|z'-z''|}^{\infty}\mathrm{d}T 
J_{0}\left(m\sqrt{T^2-|z'-z''|^2}\right)e^{-i(\omega_{\mathrm{in}}+\omega_{\bot})T}- \\
e^{-i k_{\mathrm{in}}z''}e^{i(\omega_{\mathrm{in}}-\omega_{\bot})t'} e^{i(\arctan(z''/z_r^{\mathrm{in}})-\psi_{\bot})}
\int_{|z'-z''|}^{\infty}\mathrm{d}T  J_{0}\left(m\sqrt{T^2-|z'-z''|^2}\right)e^{-i(\omega_{\mathrm{in}}-\omega_{\bot})T} \\ +c.c.\Biggr]\ . \label{eq:phi_int_int_2}
\end{multline}
The two integrals over $T$  which appear in the above equation evaluate \cite{Gradshteyn} to
\begin{multline}
\int_{|z'-z''|}^{\infty}\mathrm{d}T  J_{0}\left(m\sqrt{T^2-|z'-z''|^2}\right)e^{-i (\omega_{\mathrm{in}}\pm\omega_{\bot}) T} = \\
-\frac{i\theta(|\omega_{\mathrm{in}}\pm\omega_{\bot}|-m)\mathrm{sgn}(\omega_{\mathrm{in}}\pm\omega_{\bot})}{k_{\mathrm{ax}}^{\pm}} \\
\times e^{-i\mathrm{sgn}(\omega_{\mathrm{in}}\pm\omega_{\bot})|z'-z''|k_{\mathrm{ax}}^{\pm}}+ \frac{\theta(m-|\omega_{\mathrm{in}}\pm\omega_{\bot})|)}{k_{\mathrm{os}}^{\pm}}
e^{-|z'-z''|k_{\mathrm{os}}^{\pm}}\ ,   \label{eq:J0_int}
\end{multline}
where we have abbreviated $k_{\mathrm{ax}}^{\pm}=\sqrt{(\omega_{\mathrm{in}}\pm\omega_{\bot})^2-m^2}$ and $k_{\mathrm{os}}^{\pm}=\sqrt{m^2-(\omega_{\mathrm{in}}\pm\omega_{\bot})^2}$. 

Considering the combined $t'$ and $z'$ dependence of the axion field $\phi$ in Eqs. \eqref{eq:phi_int_int_2} and \eqref{eq:J0_int}, we can already interpret this intermediate result.
The first contribution of the integral in Eq. \eqref{eq:J0_int} encodes the situation where the axion is on shell and the frequency of the outgoing axion is equal to the sum or the
difference of the frequencies of the interacting laser beams.
The outgoing axion then propagates with $\omega_{\mathrm{ax}}^{\pm}\equiv\omega_{\mathrm{in}}\pm\omega_{\bot}$ and wave vector $k_{\mathrm{ax}}^{\pm}$.
As already suggested by physical intuition, this can only happen if $|\omega_{\mathrm{in}}\pm\omega_{\bot}|$ is larger than the mass of the axion $m$, as encoded by the
theta function in front. In addition, the axion wave carries transmitted and reflected parts, depending on the sign of $(z'-z'')$. 

The second contribution of the integral in Eq. \eqref{eq:J0_int} corresponds to the situation where the mass of the axion is larger than the sum or the difference of
$\omega_{\mathrm{in}}$ and $\omega_{\bot}$, respectively. In these situations, the axion production is off shell (os), and the axion wave decays exponentially fast with a decay
constant of $k_{\mathrm{os}}^{\pm}$.
Naturally, this is not the physical situation in which we are interested. 

Thus, specializing to $|\omega_{\mathrm{in}}\pm\omega_{\bot}|>m$ in the following, we find by combining Eqs. \eqref{eq:J0_int} and \eqref{eq:phi_int_int_2}:
\begin{multline}
\phi(z',t')=\frac{1}{8}g E_{\mathrm{in}} E_{\bot}
\Biggl[\frac{i}{k_{\mathrm{ax}}^{+}}e^{i(\omega_{\mathrm{in}}+\omega_{\bot})t'}e^{-i\mathrm{sgn}(z'-z'')k_{\mathrm{ax}}^{+}z'} e^{i\psi_{\bot}}\\
\times \int_{-\infty}^{\infty} \frac{\mathrm{d}z''}{\sqrt{1+(z''/z_r^{\mathrm{in}})^2}} e^{-(z''/w_0^{\bot})^2} e^{i(- k_{\mathrm{in}}+\mathrm{sgn}(z'-z'')k_{\mathrm{ax}}^{+})z''}
e^{i\arctan(z''/z_r^{\mathrm{in}})} \\
-\frac{i\mathrm{sgn}(\omega_{\mathrm{in}}-\omega_{\bot})}{k_{\mathrm{ax}}^{-}}e^{i(\omega_{\mathrm{in}}-\omega_{\bot})t'}
e^{-i\mathrm{sgn}(z'-z'')\mathrm{sgn}(\omega_{\mathrm{in}}-\omega_{\bot})k_{\mathrm{ax}}^{-}z'} e^{-i\psi_{\bot}}\\
\times \int_{-\infty}^{\infty} \frac{\mathrm{d}z''}{\sqrt{1+(z''/z_r^{\mathrm{in}})^2}} e^{-(z''/w_0^{\bot})^2}
e^{i(- k_{\mathrm{in}}+\mathrm{sgn}(z'-z'')\mathrm{sgn}(\omega_{\mathrm{in}}-\omega_{\bot})k_{\mathrm{ax}}^{-})z''}e^{i\arctan(z''/z_r^{\mathrm{in}})} \\
+c.c.\Biggr] \ . \label{eq:general_int}
\end{multline}
In order to evaluate the remaining integrals over $z''$, it is useful to employ the identity
\begin{equation}
\frac{e^{i \arctan{(z/z_r)}}}{\sqrt{1+(z/z_r)^2}}=\frac{1}{1- i(z/z_r)}= \int_0^{\infty} \mathrm{d}S e^{-(1-i(z/z_r))S} \ . \label{eq:arctan}
\end{equation}
In this way, the integration over $z''$ in Eq. \eqref{eq:general_int} is Gaussian and can easily be performed.
The remaining integration over $S$ is then most conveniently written in terms of the
Error function $\mathrm{erf}(x)=(2/\sqrt{\pi})\int_0^x \mathrm{d}S \exp(-S^2)$.
Eq. \eqref{eq:general_int} evaluates to 
\begin{multline}
\phi(z',t')=-\pi z_r^{\mathrm{in}}\frac{1}{4}g E_{\mathrm{in}} E_{\bot}
\Biggl[\frac{1}{k_{\mathrm{ax}}^{+}}\sin((\omega_{\mathrm{in}}+\omega_{\bot})t'-\mathrm{sgn}(z'-z'')k_{\mathrm{ax}}^{+}z'+\psi_{\bot})) \\
\times \left(1-\mathrm{erf}\left(\frac{z_r^{\mathrm{in}}}{w_0^{\bot}}+\frac{\Delta k_{\bot}^{+} w_0^{\bot}}{2}\right)\right)
\exp\left(\Delta k_{\bot}^{+} z_r^{\mathrm{in}}+ \left(\frac{z_r^{\mathrm{in}}}{w_0^{\bot}}\right)^2\right) \\
-\frac{\mathrm{sgn}(\omega_{\mathrm{in}}-\omega_{\bot})}{k_{\mathrm{ax}}^{-}}
\sin((\omega_{\mathrm{in}}-\omega_{\bot})t'-\mathrm{sgn}(z'-z'')\mathrm{sgn}(\omega_{\mathrm{in}}-\omega_{\bot})k_{\mathrm{ax}}^{-}z'-\psi_{\bot})) \\
\times \left(1-\mathrm{erf}\left(\frac{z_r^{\mathrm{in}}}{w_0^{\bot}}+\frac{\Delta k_{\bot}^{-} w_0^{\bot}}{2}\right)
\exp\left(\Delta k_{\bot}^{-} z_r^{\mathrm{in}}+ \left(\frac{z_r^{\mathrm{in}}}{w_0^{\bot}}\right)^2\right) \right)\Biggr] \ , \label{eq:general_phi}
\end{multline}
where we have defined
\begin{eqnarray}
\Delta k_{\bot}^{+}&=& -k_{\mathrm{in}}+\mathrm{sgn}(z'-z'')k_{\mathrm{ax}}^{+} \label{eq:delta_k_j_pl} \\
\Delta k_{\bot}^{-}&=& -k_{\mathrm{in}}+\mathrm{sgn}(z'-z'')\mathrm{sgn}(\omega_{\mathrm{in}}-\omega_{\bot})k_{\mathrm{ax}}^{-}\ . \label{eq:delta_k_j_min}
\end{eqnarray}
In summary, Eq. \eqref{eq:general_phi} tells us that the axion wave is composed of two partial waves with frequencies $\omega_{\mathrm{ax}}^{\pm}=(\omega_{\mathrm{in}} \pm \omega_{\bot})$,
which both have transmitted and reflected parts corresponding to $\mathrm{sgn}(z'-z'')=\pm 1$, respectively.

We find that each of these partial waves in the case of transmission and reflection is tied to a corresponding amplitude, which is a combination of an exponential and an
error function. The basic effect of this factor is that for given beam parameters the partial waves have maximal amplitude for $\Delta k_{\bot}^{\pm}\simeq 0$
if $ z_r^{\mathrm{in}}\gtrsim w_0^{\bot}$ and for $\Delta k_{\bot}^{\pm}\lesssim 0$ if $ z_r^{\mathrm{in}}\lesssim w_0^{\bot}$ and decays quickly otherwise.
The quantitative impact of this damping term, of course, depends on the absolute values of $z_r^{\mathrm{in}}$ and $w_0^{\bot}$.
As we will see later on, for experimentally feasible $z_r^{\mathrm{in}}$ and $w_0^{\bot}$, it is reasonable to assert the condition $\Delta k_{\bot}^{\pm}\simeq 0$.
In particular, we will also find an additional damping term for the back-conversion process below.

As one can conclude from Eqs. \eqref{eq:delta_k_j_pl} and \eqref{eq:delta_k_j_min}, the origin of this damping is conservation of three-momentum of the photon and the ALP:
Only if the momentum in the conversion process is conserved to a good approximation, the amplitude of the partial wave will persist undamped. Let us emphasize that the momentum of the
external beam does not enter $\Delta k_{\bot}^{\pm}$ at this point, since the external beam propagates transversal to the $z$ axis and thus there is no net transfer of momentum in
the $z$ direction. This will become important later on.
Note also that this damping factor in practice determines the sensitivity to the mass of the axion which is contained in $k_{\mathrm{ax}}^{\pm}$. I.e. given two laser frequencies,
efficient conversion into axions can only happen, if the mass of the axion is next to resonance, such that the sum of momenta vanishes approximately\footnote{The above
conversion properties are
reminiscent of the processes of sum-frequency and difference-frequency generation,
known from Nonlinear Optics, see e.g. \cite{boyd}. For these processes, a suitable medium with nonlinear dielectric permittivity is used to produce
light beams whose frequency is equal to the sum or
difference of the frequencies of the input beams. There, an efficient conversion can only happen, if so called phase-matching conditions are fulfilled.
These phase matching conditions are analogous to the above condition of three-momentum conservation.}: $\Delta k_{\bot}^{\pm}\simeq 0$.
Since $\Delta k_{\bot}^{\pm}$ depends linearly on the ALP mass, there is exactly one resonant
ALP mass for given frequencies $\omega_{\mathrm{in}}$ and $\omega_{\bot}$.

Before we proceed with the calculation of the back-conversion of the ALPs into photons, let us determine the resonant ALP masses $m$ in the conversion process.
In vacuum ($k=\omega$), we have the requirements
\begin{eqnarray}
\Delta k_{\bot}^{+}&=& -\omega_{\mathrm{in}}+\mathrm{sgn}(z'-z'')\sqrt{(\omega_{\mathrm{in}}+\omega_{\bot})^2-m^2} \stackrel{!}{\simeq} 0 \label{eq:k_j_plus} \\
\Delta k_{\bot}^{-}&=& -\omega_{\mathrm{in}}+\mathrm{sgn}(z'-z'')\mathrm{sgn}(\omega_{\mathrm{in}}-\omega_{\bot})\sqrt{(\omega_{\mathrm{in}}-\omega_{\bot})^2-m^2}
\stackrel{!}{\simeq} 0 \ . \label{eq:k_j_minus}
\end{eqnarray}
Eq. \eqref{eq:k_j_plus} is solved in the case of transmission (i.e. $\mathrm{sgn}(z'-z'')=+1$) by choosing $m= \sqrt{\omega_{\bot}^2+2\omega_{\mathrm{in}}\omega_{\bot}}$.
For \eqref{eq:k_j_minus}, there is in principle the resonant solution $m = \sqrt{\omega_{\bot}^2-2\omega_{\mathrm{in}}\omega_{\bot}}$ which
implies $\frac{1}{2}\omega_{\bot}>\omega_{\mathrm{in}}$  for positive axion masses implying reflection  (i.e. $\mathrm{sgn}(z'-z'')=-1$) but requiring also a 
negative value for $\omega_{\mathrm{ax}}^{-}=\omega_{\mathrm{in}}-\omega_{\bot}$.
Thus, the latter is an unphysical solution.

Let us summarize our findings from Eq. \eqref{eq:general_phi}: induced by the interaction of the probe field $\omega_{\mathrm{in}}$ with the external beam $\omega_{\bot}$,
one obtains transmitted and reflected axion waves with frequencies  $\omega_{\mathrm{ax}}=\omega_{\mathrm{in}}\pm\omega_{\bot}$.
For our purposes, we focus on the transmitted partial wave with frequency $\omega_{\mathrm{ax}}=\omega_{\mathrm{in}}+\omega_{\bot}$ 
since it acquires an undamped amplitude for ALP masses $m$
which are close to a resonant mass
\begin{equation} 
m_{\bot} = \sqrt{\omega_{\bot}^2+2\omega_{\mathrm{in}}\omega_{\bot}} \label{eq:m_bot} \ .
\end{equation}
Thus, for the following calculation of the reconversion process, we employ for clarity only the transmitted axion wave with frequency $\omega_{\mathrm{ax}}^{+}$:
\begin{multline}
\phi^{(\mathrm{T})}(z',t')\approx-\pi z_r^{\mathrm{in}}\frac{1}{4}g E_{\mathrm{in}} E_{\bot}
\Biggl[\frac{1}{k_{\mathrm{ax}}^{+}}\sin((\omega_{\mathrm{in}}+\omega_{\bot})t'-k_{\mathrm{ax}}^{+}z'+\psi_{\bot}))\\
\left(1-\mathrm{erf}\left(\frac{z_r^{\mathrm{in}}}{w_0^{\bot}}+\frac{\Delta k_{\bot}^{+} w_0^{\bot}}{2}\right)\right)\exp\left(\Delta k_{\bot}^{+} z_r^{\mathrm{in}}+
\left(\frac{z_r^{\mathrm{in}}}{w_0^{\bot}}\right)^2\right)\Biggr] \ .
\label{eq:phi_T_plus}
\end{multline}
Let us finally remark that the damping factor of the amplitude encountered above is in fact not an artifact of the Gaussian beam form.
This can easily be checked by omitting all factors containing $w_0^{\bot}$ and $z_r^{\mathrm{in}}$ in Eq. \eqref{eq:general_int}, which amounts to calculating the
interaction between two plain waves. Then, integrating over $z''$ over a length $L$  of the interaction region,
one finds that the amplitudes of the axion partial waves are proportional to a factor of $\sin(\Delta k_{\bot}^{\pm} L/2)/(\Delta k_{\bot}^{\pm})$, respectively.
Hence, also for a plain wave approximation, the amplitudes are peaked around $\Delta k_{\bot}^{\pm}\simeq 0$.

From this ansatz, the conversion amplitude for the ALP in a temporally and
spatially constant external field follows in the limit
$\omega_{\bot},k_{\bot}\rightarrow 0$.  The square of this amplitude is given
in Eq. \eqref{eq:P_constant_ext} in the limit\footnote{Note that in this situation for
  large axion masses $m\simeq \omega_{\mathrm{in}}$, the computation of the
  conversion probabilities requires great care, as discussed in detail in
  \cite{Adler:2008gk}. However, generically, the most stringent exclusion
  bounds on ALPs for constant external fields are obtained for masses $m\ll
  \omega_{\mathrm{in}}$. In the end, we will compare the discovery potential
  for the purely-laser based setup to these bounds.}  $m\ll
\omega_{\mathrm{in}}$, and will be used later on for a qualitative comparison
to dipole LSW experiments.

\section{Axion-Photon-Conversion \label{sec:ax_pho_con}}

We now turn to the back-conversion of the ALPs into photons, by virtue of Eq. \eqref{eq:photon_eom_1d}.
With the Green's function in Eq. \eqref{eq:G0} and using  $e_{\mathrm{out}}=-\partial_{t}a_{\mathrm{out}}$, we have to evaluate
\begin{multline}
e_{\mathrm{out}}(z,t) = \frac{1}{2} g \int_{-\infty}^{\infty} \mathrm{d}z' \int_{-\infty}^{\infty} \mathrm{d}t' \delta\left((t-t')-|z-z'|\right) \\
\left[[B_k^{y}(z',t') \partial_{t'} \phi(z',t')]+ E_k^{x}(z',t') \partial_{z'} \phi(z',t')]\right] \ . \label{eq:eout}
\end{multline}
As in the previous section we choose the external field for the back-conversion with hindsight: Assuming propagation of the electromagnetic wave along the negative $z$ axis,
i.e. counter-propagating to the transmitted axion wave, we set
\begin{eqnarray}
E_{k}^{x}(z,t)&=& \frac{E_{\parallel}}{\sqrt{1+\left(z/z_r^{\parallel}\right)^2}}
\sin\left(\omega_{\parallel} t + k_{\parallel} z-\arctan\left(\frac{z}{z_r^{\parallel}}\right) + \psi_{\parallel}\right) \label{eq:E_k_field} \\
B_{k}^{y}(z,t)&=& -E_{\parallel}^{x}(z,t)=- \frac{E_{\parallel}}{\sqrt{1+\left(z/z_r^{\parallel}\right)^2}}
\sin\left(\omega_{\parallel} t + k_{\parallel} z -\arctan\left(\frac{z}{z_r^{\parallel}}\right) + \psi_{\parallel}\right)\ , \label{eq:B_k_field}
\end{eqnarray}
with beam parameter definitions as above.

In the following it is convenient to use the axion and laser wave again in complex notation. Plugging the approximate transmitted axion wave
from Eq. \eqref{eq:phi_T_plus} and the external beam
(Eqs. \eqref{eq:E_k_field} and \eqref{eq:B_k_field}) into Eq. \eqref{eq:eout},
evaluating  the derivatives acting on $\phi(z,t)$ and 
integration over $t'$, we find
\begin{multline}
e_{\mathrm{out}}(z,t) = \frac{1}{32} g^2 \pi z_r^{\mathrm{in}} E_{\mathrm{in}} E_{\bot} E_{\parallel} \\
\times
\left(1-\mathrm{erf}\left(\frac{z_r^{\mathrm{in}}}{w_0^{\bot}}+\frac{\Delta k_{\bot}^{+} w_0^{\bot}}{2}\right)\right)
\exp\left(\Delta k_{\bot}^{+} z_r^{\mathrm{in}}+ \left(\frac{z_r^{\mathrm{in}}}{w_0^{\bot}}\right)^2\right) \left(\frac{\omega_{\mathrm{in}}+\omega_{\bot}}{k_{\mathrm{ax}}^{+}}+1\right)\\
\Biggl[\frac{1}{i} e^{i (\omega_{\mathrm{in}}+\omega_{\bot}+\omega_{\parallel})(t-\mathrm{sgn}(z-z')z)}
e^{i(\psi_{\bot}+\psi_{\parallel})}\int_{-\infty}^{\infty} \mathrm{d}z' \frac{1}{\sqrt{1+(z'/z_r^{\parallel})^2}} e^{i \Delta k_{\parallel}^{+} z'}
e^{-i\arctan\left(\frac{z'}{z_r^{\parallel}}\right)}\\-
\frac{1}{i} e^{i (\omega_{\mathrm{in}}+\omega_{\bot}-\omega_{\parallel})(t-\mathrm{sgn}(z-z')z)}e^{i(\psi_{\bot}-\psi_{\parallel})}\int_{-\infty}^{\infty}
\mathrm{d}z' \frac{1}{\sqrt{1+(z'/z_r^{\parallel})^2}} e^{i \Delta k_{\parallel}^{-} z'} e^{i\arctan\left(\frac{z'}{z_r^{\parallel}}\right)} +c.c.\Biggr]\ . \label{eq:e_out_int}
\end{multline}
Here we have defined 
\begin{eqnarray}
\Delta k_{\parallel}^{+}&=& -k_{\mathrm{ax}}^{+}+k_{\parallel} + \mathrm{sgn}(z-z') (\omega_{\mathrm{in}}+\omega_{\bot}+\omega_{\parallel}) \label{eq:delta_k_pl}\\
\Delta k_{\parallel}^{-}&=& -k_{\mathrm{ax}}^{+}-k_{\parallel} + \mathrm{sgn}(z-z') (\omega_{\mathrm{in}}+\omega_{\bot}-\omega_{\parallel})\ . \label{eq:delta_k_min}
\end{eqnarray}
Eq. \eqref{eq:e_out_int} resembles the situation of the ALP \textit{production}, as the outgoing electromagnetic wave $e_{\mathrm{out}}$ essentially consists of two partial waves
with frequencies $\omega_{\mathrm{out}}^{\pm} =\omega_{\mathrm{in}}+\omega_\bot \pm\omega_\parallel$. Again, each partial wave has a transmitted ($\mathrm{sgn}(z-z')=1$) and a
reflected ($\mathrm{sgn}(z-z')=-1$) contribution.
In order to determine the corresponding amplitudes of the partial waves, it is necessary to perform the remaining integration over $z'$.
To this end, we make use of the first identity in \eqref{eq:arctan} and perform the  spatial integration over a closed contour in the complex $z'$ plane.
In this manner, we find for the integrals in Eq. \eqref{eq:e_out_int}
\begin{multline}
\int_{-\infty}^{\infty} \mathrm{d}z' \frac{1}{\sqrt{1+\left(z'/z_r^\parallel\right)^2}} e^{\mp i\arctan\left(\frac{z'}{z_r^\parallel}\right)} e^{i \Delta k_\parallel^{\pm} z'}= \\
\int_{-\infty}^{\infty} \mathrm{d}z' \frac{1}{1\pm i\left(z'/z_r^\parallel\right)}e^{i \Delta k_\parallel^{\pm} z'}= \pi z_r^\parallel (1\pm \mathrm{sgn}(\Delta k_\parallel^{\pm}))
e^{-z_r^\parallel|\Delta k_\parallel^{\pm}|} \ . \label{eq:backconv_ampl}
\end{multline}
Analogous to the previous section, the partial wave amplitudes in Eq. \eqref{eq:backconv_ampl} entering $e_{\mathrm{out}}$ are strongly peaked at vanishing $\Delta k_\parallel^{\pm}$.
As in the photon-ALP conversion, this peak structure can be understood in terms of three-momentum conservation, cf. Eqs. \eqref{eq:delta_k_pl} and \eqref{eq:delta_k_min}, where the last
term incorporates the momentum of the outgoing photon.

In addition, sharp cutoffs arise for $\Delta k_\parallel^{+}<0$ and $\Delta k_\parallel^{-}>0$ in Eq. \eqref{eq:backconv_ampl} through the signum function, respectively.
Physically, this is due to the Gouy phase anomaly \cite{Gouy}; this behavior is also well known in the context of
nonlinear interactions with focused Gaussian beams in media, see e.g. \cite{boyd}. However, one can check by numerical integration that the sharp cutoff of the signum function
is in fact washed out for integrations over finite interaction regions. Thus, for finite, physical interaction regions, the amplitudes are maximized for $\Delta k_\parallel^{\pm}\simeq 0$.

Before we continue with the evaluation of $e_{\mathrm{out}}$, let us check the \textit{compatibility} of the three-momentum conservations for the two conversion processes.
Only if the conservation of three-momentum is obeyed at both conversions at the same time, the overall amplitude is undamped.
In addition to conservation of three-momentum, we want to make the important additional
requirement that $\omega_{\mathrm{out}}\neq \omega_{\mathrm{in}}$, as signature for the conversion processes to have taken place at all.

To this end, we consider Eqs. \eqref{eq:delta_k_pl} and \eqref{eq:delta_k_min} again in vacuum (i.e. $\omega=k$). In order to compare with the conservation of momentum
in the photon-axion conversion process encoded in $\Delta k_\bot^{+}$ (see Eq. \eqref{eq:k_j_plus}), we multiply $\Delta k_\parallel^{\pm}$ by $-1$. This is justified, since the
exponential damping depends only on the modulus of $\Delta k_\parallel^{\pm}$. Eqs. \eqref{eq:delta_k_pl} and \eqref{eq:delta_k_min} then read:
\begin{eqnarray}
-\Delta k_\parallel^{+}&=& \sqrt{(\omega_{\mathrm{in}}+\omega_\bot)^2-m^2} -
\mathrm{sgn}(z-z') (\omega_{\mathrm{in}}+\omega_\bot+\omega_\parallel)-\omega_\parallel \stackrel{!}{\simeq} 0 \label{eq:delta_k_pl_vac}\\
-\Delta k_\parallel^{-}&=& \sqrt{(\omega_{\mathrm{in}}+\omega_\bot)^2-m^2} -
\mathrm{sgn}(z-z') (\omega_{\mathrm{in}}+\omega_\bot-\omega_\parallel)+\omega_\parallel \stackrel{!}{\simeq} 0 \ . \label{eq:delta_k_min_vac}
\end{eqnarray}
By comparing the above conditions for $\Delta k_\parallel^{\pm}$ with that for
$\Delta k_\bot^{+}$ in Eq. \eqref{eq:k_j_plus}, we notice that we need the
transmitted part of the outgoing wave, corresponding to $\mathrm{sgn}(z-z')
=+1$ also for the back-conversion process. Otherwise the required relative sign
between $\omega_{\mathrm{in}}$ and the axion wave vector in order to
satisfy \Eqref{eq:k_j_plus} can not be reproduced.

In order to fulfill the condition for $\Delta k_\parallel^{+}$ in Eq. \eqref{eq:delta_k_pl_vac} and the condition for $\Delta k_\bot^{+}$ at the same time,
a negative frequency contribution of either $\omega_\bot$ or $\omega_\parallel$ would then be needed which is clearly unphysical. For this reason we drop this option in the following.

By contrast, it can be seen that it is possible to obey momentum conservation for the conversion and back-conversion processes simultaneously via Eq. \eqref{eq:delta_k_min_vac}:
By setting $\omega_\bot=2 \omega_\parallel$, we see that $\Delta k_\parallel^{-}=\Delta k_\bot^{+}=0$ if the axion mass satisfies Eq. \eqref{eq:m_bot}. 
In addition, the outgoing electromagnetic wave in this situation has a frequency $\omega_{\mathrm{out}}=\omega_{\mathrm{in}}+\frac{1}{2} \omega_\bot=\omega_{\mathrm{in}}+\omega_\parallel$
which is different from the frequency of the incoming wave. We conclude that also our second requirement for $e_{\mathrm{out}}$ is met.

At this point, it also becomes clear why it is crucial for the frequency shift of the outgoing wave to arise that the momentum of the first external laser beam  does not enter the requirement
for $\Delta k_\bot^{+}$ in Eq. \eqref{eq:delta_k_j_pl}: Depending on the relative signs of the momenta in the two conversion processes, conservation of momentum 
could only be achieved in both processes for $\omega_\bot=\omega_\parallel$.
However, this would immediately imply
$\omega_{\mathrm{in}}=\omega_{\mathrm{out}}$ being problematic for
experimental observation and require again an LSW setup which is difficult to
conceive for high-intensity fields.

In total, it is the above ''mismatch'' of momentum conservation (as encoded in $\Delta k_{\bot/\parallel}$) and energy conservation
(as encoded in the difference $\omega_{\mathrm{out}}-\omega_{\mathrm{in}}$), which is essential for a net transfer of energy to the outgoing photon, favoring experimental detectability.

To summarize, it is the transmitted (T) contribution of the second partial wave in Eq. \eqref{eq:e_out_int} for which both respective momentum conservation conditions can be fulfilled 
simultaneously 
and a frequency shift with respect to the incoming wave arises. 
Combining Eqs.  \eqref{eq:e_out_int}
and Eq. \eqref{eq:backconv_ampl} and substituting $\omega_\bot=2 \omega_\parallel$, we focus for the remainder on
\begin{multline}
e_{\mathrm{out}}^{(\mathrm{T})}(z,t) \approx -\frac{1}{16} g^2 \pi^2 z_r^{\mathrm{in}} z_r^\parallel  E_{\mathrm{in}} E_\bot E_\parallel \left(1-\mathrm{erf}
\left(\frac{z_r^{\mathrm{in}}}{w_0^\bot}+
\frac{\Delta k w_0^\bot}{2}\right)\right)\exp\left(\Delta k z_r^{\mathrm{in}}+ \left(\frac{z_r^{\mathrm{in}}}{w_0^\bot}\right)^2\right) \\
\times \left(1- \mathrm{sgn}\left(\Delta k\right)\right)e^{-z_r^\parallel|\Delta k|}
\left(\frac{\omega_{\mathrm{in}}+ 2 \omega_\parallel}{k_{\mathrm{ax}}^{+}}+1\right) \sin((\omega_{\mathrm{in}}+\omega_\parallel)(t-z)+\psi_\bot-\psi_\parallel) \ , \label{eq:e_out_bot_par}
\end{multline}
where we have set $\Delta k_\bot^{+}=\Delta k_\parallel^{-} \equiv \Delta k$.

In practice it is of course not directly experimentally assessable which of
the two external beams mediates conversion and which one back-conversion. So
far, we have assumed conversion
to be induced by the field with $\bot$ orientation, and back-conversion due to the counter-propagating $\parallel$ field.
The result for the outgoing wave, however, will depend on the order of interaction. As demonstrated in the Appendix, the conversion process with the opposite order 
$\bot \leftrightarrow \parallel$ results in an
outgoing wave with frequency
$\omega_{\mathrm{out}}=\omega_{\mathrm{in}}-\omega_\parallel$, if one chooses $\omega_\bot=2\omega_\parallel$.
The resonant mass satisfying momentum conservation in both conversion processes is then given by $m_\parallel=2\sqrt{\omega_{\mathrm{in}}\omega_\parallel}$, being different
from the resonant mass at interchanged interaction order. In consequence, it is possible to probe the axion coupling space around two resonant masses within one setup.

Our calculations suggest the following experimental setup. 
Probe photons, which traverse one counter-propagating and one perpendicularly propagating laser field
with frequencies $\omega_{\parallel}$ and $\omega_{\bot}$, respectively, can experience a frequency shift due to ALP-photon mixing. This happens if the frequency 
of the external perpendicular laser has twice
the frequency of the external counter-propagating laser: $2\omega_{\parallel}=\omega_{\bot}$. The requirement for this process is the existence of ALPs with masses close to one of the
two resonant masses, which are a function of the involved laser frequencies.

These resonant masses are of the same order of magnitude as the involved frequency scales of the lasers, which in optical setups corresponds to $\sim \mathcal{O}(\mathrm{eV})$.
As this mass regime is so far largely unexplored in laboratory ALP-searches, the proposed experimental setup can be \textit{complementary} 
to the search involving dipole magnets, see below.  

It is worth emphasizing that the required frequency ratio for the two external lasers is fact an enormous experimental advantage, since it implies that indeed
only \textit{one} high-intensity laser is needed as external field, since frequency doubling \cite{shg} is a standard technique even for high-intensity lasers.
In addition, the corresponding beam parameters of the frequency doubled beam, such as the focal area can in principle be tuned independently by the use of appropriate lens systems.

\section{Exclusion limits\label{sec:results}}

Let us explore the parameter range in the ALP mass and coupling plane which can be probed within the presented setup.

The number of photons in the beams is proportional to the square of the field amplitudes, being a function of time.
However, as the pulse lengths in consideration imply a large number of wave trains,
a good approximation of the number of frequency shifted outgoing photons $N_{\mathrm{out}}$ as a function of the number of incoming photons $N_{\mathrm{in}}$
can be read off from a comparison of Eq. \eqref{eq:e_out_bot_par} (or \eqref{eq:e_out_par_bot_app}) with the incoming field $e_{\mathrm{in}}$ in
Eq. \eqref{eq:probe_model}. 
\begin{equation}
N_{\mathrm{out}}(\omega_{\mathrm{in}}\pm\omega_\parallel) \simeq N_{\mathrm{in}}(\omega_{\mathrm{in}})  N_{\mathrm{shot}} \alpha_{\pm}^2 \label{eq:Ninout} \ .
\end{equation}
The parameter $N_{\mathrm{shot}}$ counts the number of laser shots used for a measurement. It is determined by the total measurement time of data accumulation times the repetition
rate of the lasers.
The quantity $\alpha_{\pm}^2$ is a measure for the probability of the photon-axion-photon conversion. Here, $\alpha_{+}$ denotes the conversion amplitude for outgoing photons
of frequency $\omega_{\mathrm{out}}=\omega_{\mathrm{in}}+\omega_{\parallel}$ whereas $\alpha_{-}$ is the conversion amplitude for photons with
frequency $\omega_{\mathrm{out}}=\omega_{\mathrm{in}}-\omega_{\parallel}$. They read:
\begin{eqnarray}
\alpha_{+} &=& - \frac{1}{16} g^2 \pi^2 z_r^{\mathrm{in}} z_r^{\parallel} E_{\bot} E_{\parallel}
\left(1-\mathrm{erf}\left(\frac{z_r^{\mathrm{in}}}{w_0^{\bot}}+\frac{\Delta k w_0^{\bot}}{2}\right)\right)
\left(\frac{\omega_{\mathrm{in}}+ 2 \omega_{\parallel}}{\sqrt{(\omega_{\mathrm{in}}+2 \omega_{\parallel})^2-m^2}}+1\right) \nonumber \\
&{}& \times \exp\left(\Delta k z_r^{\mathrm{in}}+ \left(\frac{z_r^{\mathrm{in}}}{w_0^{\bot}}\right)^2\right) 
(1- \mathrm{sgn}(\Delta k))e^{-z_r^{\parallel}|\Delta k|}\ , \\
\alpha_{-} &=& \frac{1}{8} g^2  \pi^{3/2} \frac{z_r^{\mathrm{in}}z_r^\parallel}{z_r^{\mathrm{in}}+z_r^\parallel} w_0^{\bot} E_{\parallel} E_{\bot}
\frac{\omega_{\mathrm{in}}+\omega_{\parallel}}{\sqrt{(\omega_{\mathrm{in}}+ \omega_{\parallel})^2-m^2}} \nonumber \\
&{}& \times \left[(1-\mathrm{sgn}(\delta k)) e^{\delta k z_r^{\mathrm{in}}}+(1+\mathrm{sgn}(\delta k )) e^{-\delta k z_r^{\parallel}}\right] 
e^{-\frac{1}{4} (w_0^{\bot} \delta k)^2} \label{eq:alpha_pl} \ , 
\end{eqnarray}
where we have inserted the respective axion wave vectors for clarity.
The parameters $\Delta k$ and $\delta k$ reduce to
\begin{eqnarray}
\Delta k &=& -\omega_{\mathrm{in}}+\sqrt{(\omega_{\mathrm{in}}+2 \omega_{\parallel})^2-m^2}\ , \\
\delta k &=& -\omega_{\mathrm{in}}+\omega_{\parallel}+\sqrt{(\omega_{\mathrm{in}}+ \omega_{\parallel})^2-m^2} \ .
\end{eqnarray}
As already denoted above, in order to experimentally assess the induced frequency shift,
it is sufficient to have just one high-intensity laser as external field in this configuration.
One part of the external beam has to be frequency doubled, such that $\omega_{\bot}=2 \omega_{\parallel}$, while a delay line must ensure
the simultaneous overlap of the focal spots of the fundamental and the frequency doubled beam mode as well as the focal spot of the probe beam.

In addition, we demand that the three frequency modes satisfy $\omega_{\mathrm{in}}+\omega_{\mathrm{\parallel}} \nsim \omega_{\mathrm{\bot}}$, in order to make 
the frequency shifted photons at  $\omega_{\mathrm{out}}$  distinguishable from the frequency components already used in the process.  We will detail on this requirement below.

Further, it appears advisable in practice to realize the setup with a slight deviation from the exact orthogonal geometry in order
to facilitate the detection of the frequency shifted photons off the main optical axis in order to reduce the noise amplitude\footnote{For a quantitative discussion
of a non-orthogonal setup, the above calculation, of course, has to be extended to three spatial dimensions.
Qualitatively, we expect that, for deviations from the orthogonal geometry, the resonance conditions $\Delta k\simeq 0$ and $\delta k \simeq 0$
receive an angular dependence based on the three-momentum transfer in all spatial directions. In consequence, the necessary condition
$\omega_{\bot}=2 \omega_{\parallel}$ is expected be modified as well. However, as this relation is experimentally easily accessible by frequency doubling,
a beam geometry close to the orthogonal setup appears to remain viable.}.

Let us now determine the discovery potential in the ALP mass-coupling plane for an operational high-intensity facility. 
First, we discuss a possible setup at Jena \cite{IOQ,HIJ}. In the near future, the Multi-Terawatt class laser JETI and the Petawatt class laser
POLARIS \cite{POLARIS} can be focused simultaneously and synchronized into a single target chamber.
Thus we can employ the lower-intensity laser JETI for providing the probe photons, while POLARIS is used
to create the two external fields for the conversion processes.

Let us first consider the focal parameters of the two lasers.
 
To achieve maximum field strengths for POLARIS and a good bunching of the probe photons of JETI, we need very small focal spots on the order of
the diffraction limit.
To be more precise, if we define the effective diameter of the focal spot to contain $86\%$ (corresponding to the $1/e^2$-criterion) of the focused beam energy, see below,
one has the estimate \cite{Siegman} $w_0 \simeq f^{\#} \lambda$, with the so-called $f$-number $f^{\#}$ of the focusing lens, which characterizes the ratio of the focal length and the focusing
aperture diameter.
Ambitious, but feasible values for the $f$-number can be as low as $f^{\#}=1$.
Recall that the corresponding Rayleigh length is obtained from the waist size through $z_r=\frac{\pi w_0^2}{\lambda}$.

The laser system POLARIS is designed to provide a peak power of around $P=1 \mathrm{PW}$
(attained through $150\mathrm{J}$ at pulse lengths $\tau_{\mathrm{ext}}\simeq 150\mathrm{fs}$), optimized for a 
central wavelength of $\lambda_{\mathrm{ext}}= 1035\mathrm{nm}$, corresponding
to $\omega_\parallel=1.20 \mathrm{eV}$ and thus $\omega_\bot=2.40 \mathrm{eV}$. In consequence, we find 
$w_0^{\bot}\simeq 5 \frac{\lambda_{\mathrm{ext}}}{2} \approx 13.1 \mathrm{eV^{-1}}$, for the frequency doubled beam
where we have chosen an $f$-number $f^{\#}=5$ in order to fit the probe photons into the external field,
see below.

Further, we estimate the Rayleigh length for the counter-propagating fundamental beam as  $z_r^{\parallel}\simeq \pi (f^{\#})^2 \lambda_{\mathrm{ext}} \simeq 16.4 \mathrm{eV^{-1}}$
for an $f$-number of $f^{\#}=1$.

The obtained intensities within the focal spot yield
$I_\parallel= 0.86 \frac{1}{2}\frac{P}{A_\parallel} = 2.05\times 10^{16}\mathrm{eV}^4$
and
$I_\bot= 0.86 \frac{1}{2}\frac{P}{A_\bot} =  3.28\times 10^{15}\mathrm{eV}^4$, where the reduction of intensity for the $\bot$ beam through losses in
the frequency doubling process has yet to be accounted for.

Here, $A_{(\bot/\parallel)}=\left(w_0^{(\bot/\parallel)}\right)^2 \pi$ is the area of the focal spot and the
factor of $1/2$ enters due to the splitting of POLARIS into two separate beams $\omega_{\parallel}$ and $\omega_{\bot}$.
It is clear, however, that these are upper theoretical estimates for the achievable intensities which will be certainly modified by the circumstances of the experimental setup.

From these intensities, the peak electric field strength for the fundamental mode is $E_\parallel=\sqrt{I_\parallel} \simeq 1.43\times 10^8 \mathrm{eV}^2$.
Further , it is appropriate to assume a relatively moderate conversion efficiency of $40\% $ for the frequency doubled beam at these field strengths, yielding
$E_\bot=\sqrt{0.4 I_\bot} \approx 3.62\times 10^7 \mathrm{eV}^2$.

For high intensities, the external pulses as well as the probe beam must not only be spatially but also temporally well focused. 
Naturally, the pulses then have a spectral width $\Delta \omega$, which, for Gaussian pulses is related to the pulse length as $\Delta \omega\simeq 0.44 \frac{2\pi}{\tau}$.

In order to detect the frequency shifted photons at $\omega_{\mathrm{out}}=\omega_{\mathrm{in}}\pm\omega_{\parallel}$ with low noise,
these photons should lie well outside the spectral widths $\Delta \omega_{\mathrm{in}}$,
$\Delta \omega_{\mathrm{\parallel}}$ and  $\Delta \omega_{\mathrm{\bot}}$, centered around $\omega_{\mathrm{in}}$ and $\omega_{\mathrm{\parallel}}$ and 
$\omega_{\mathrm{\bot}}$, respectively.

As the external pulses are comparatively long, we obtain a small 
spectral width
of $\Delta \omega_{\mathrm{ext}}\simeq 0.01 \mathrm{eV}$. With the above pulse length, we see that  $\tau_{\mathrm{ext}}\gtrsim z_r^{\mathrm{ext}}$ as well as
$\tau_{\mathrm{ext}}\gtrsim w_0^{\mathrm{ext}}$ are well obeyed if we assume similar focal properties for the fundamental beam mode and the frequency doubled beam.

After the second upgrade, JETI++ is expected to provide an energy of around $\mathcal{E}=3 \mathrm{J}$ per shot at a central
wavelength of around $\lambda_{\mathrm{in}}= 800\mathrm{nm}$ ($\omega_{\mathrm{in}}=1.55 \mathrm{eV}$) with pulse lengths as small as $\tau_{\mathrm{in}}\simeq 30 \mathrm{fs}$. 
Assuming $f^{\#}=1$, the smallest Rayleigh length for the probe beam is therefore
$z_r^{\mathrm{in}}\approx 12.7 \mathrm{eV}$.
From the pulse energy, the number of incoming photons per shot is given by $N_{\mathrm{in}}= \mathcal{E}/\omega_{\mathrm{in}} \approx 1.21 \times 10^{19}$.
In addition, the requirements 
$\omega_{\mathrm{out}}\notin \Delta \omega_{\mathrm{\parallel}}, \Delta\omega_{\mathrm{\bot}}, \Delta\omega_{\mathrm{\parallel}}$ are well obeyed, since the spectral width is
only $\Delta \omega_{\mathrm{in}}\simeq 0.06 \mathrm{eV}$.

As $\tau_{\mathrm{in}}\ll \tau_{\mathrm{ext}}$, all $N_{\mathrm{in}}$ JETI photons are available for the conversion process as long as
$z_r^{\mathrm{in}} \lesssim z_r^{\parallel}, w_0^{\bot}$, which is implemented above by the choice of the focusing geometry, see also Fig. \ref{fig:setup}.

A decisive experimental parameter is the pulse repetition rate which determines  $N_{\mathrm{shot}}$ for a given measurement time.
As the computation of the necessary statistics for the photon detection requires detailed knowledge of the laser specifications and the setup we perform our estimates
for $N_{\mathrm{shot}}=N_{\mathrm{out}}=1$. In particular, nonlinear processes within the experimental setup tend to modify the idealized Gaussian frequency spectra of the laser beams.
If larger statistics for the outgoing photons are required, this can always be accommodated by a larger number of shots.
In the present example, POLARIS, due to its higher energy, has the smaller repetition rate of both lasers, which is expected to approach $f_{\mathrm{rep}}\simeq 0.1 \mathrm{Hz}$.
In practice, $\mathcal{O}(100)$ shots per day can be achieved, being a huge accomplishment for a Petawatt-class laser. 
In the future, improved cooling schemes for the amplifying medium may even
lead to a further enhancement of the repetition rate.

For these parameters, the discovery potential follows from Eqs. (\ref{eq:Ninout}-\ref{eq:alpha_pl}). In Fig. \ref{fig:bounds}, we obtain two black wedge-like
curves around the resonant masses $m_\parallel=2.73 \mathrm{eV}$ and  $m_{\bot}= 3.63 \mathrm{eV}$, which are determined by the photon energies of POLARIS and JETI.
The peaks of the wedges and thus the minimal accessible coupling strength lie at $g\approx 7.7\times 10^{-6} \mathrm{GeV}^{-1}$ and 
$g\approx 1.7\times 10^{-5} \mathrm{GeV}^{-1}$, respectively.

The currently best laboratory limits on ALPs by the LSW setup of the ALPS collaboration \cite{Ehret:2010mh},
are indicated as blue-shaded area in Fig. \ref{fig:bounds}, while the
best limits on solar axions are provided by the CAST experiment \cite{CAST}, denoted by a green-dashed line. Although astrophysical considerations currently give the strongest
constraints, they are somewhat model dependent due to a different momentum-transfer regime \cite{Jaeckel:2006xm}.

In dipole experiments, the external magnetic field $B$ is essentially constant, and the axion-photon conversion and reconversion probability in vacuum are well approximated by
\begin{equation}
P_{\gamma \rightarrow \phi, \phi  \rightarrow \gamma}= \left(\frac{g B L }{2}\right)^2 \frac{\sin^2(x)}{x^2} \label{eq:P_constant_ext}\ ,
\end{equation}
for the ALP parameter space under consideration, cf. Fig. \ref{fig:bounds}.
Above, we have abbreviated  $x= m^2 L/(4 \omega_{\phi/\gamma})$, where $L$ denotes the length of the dipole magnet and $\omega_{\phi/\gamma}$ is the axion 
and photon energy, respectively.
From Eq. \eqref{eq:P_constant_ext} it is obvious that the best exclusion bounds are obtained for small arguments $x\ll 1$, i.e. for small axion masses at fixed
energies $\omega_{\phi/\gamma}$ and dipole length $L$, cf. Fig. \ref{fig:bounds}. If the axion masses become too large, the conversion probability suffers from a $x^{-2}$ suppression.

As the CAST-experiment utilizes $\omega_{\phi}\sim \mathrm{keV}$ solar axions, the best bounds are obtained for masses below $m\lesssim 1 \mathrm{eV}$. By contrast, in laboratory experiments
generically $\omega_{\gamma}\sim \mathrm{eV}$, and thus the drop-off sets in at even lower masses $m \sim 10^{-3} \mathrm{eV}$.
In both situations, the accessible mass regions can be slightly extended to higher masses by the use of buffer gas. However, the generic form of the conversion probability
for constant external fields, see Eq. \eqref{eq:P_constant_ext}, disfavors dipole searches for the exploration of higher ALP mass ranges.

For both dipole experiments, the exploration of the $m \simeq 1 \mathrm{eV}$
region is difficult and thus the purely laser-based search can complement
the existing ALP
searches in the large mass region as detailed above. An additional feature of the purely laser-based search is the strong sensitivity to the axion mass which originates from the condition 
of momentum conservation.
On the one hand this can be advantageous, since it allows for a direct estimate of the axion mass if a signal is detected.
On the other hand, it would of course be desirable to explore a larger range of the axion-mass-coupling plane within a single setup.

For this purpose, it would be favorable if the involved laser frequencies were tunable 
within a certain frequency range. 

In fact this can be realized by the use of optical parametric amplifiers (OPAs), which are employed to tune the 
frequency of an optical pump laser over a wide frequency range while retaining the temporal structure of the pulse to a good approximation. However, for today's OPAs,
the pump energy
is limited to approximately $1\mathrm{mJ}$. Since this requirement strongly limits the available intensity,
we choose to tune the frequency of the probe beam rather than the external beam which enters twice in the setup in its fundamental and frequency doubled mode.
For optical probe lasers, we have thus a limitation of the number of incoming photons to around  $N_{\mathrm{in}}\simeq10^{15}$.
Yet, as  $N_{\mathrm{in}}\sim g^4$,
cf. Eqs. (\ref{eq:Ninout}-\ref{eq:alpha_pl}), for the above considered setup
the sensitivity to the  coupling is reduced by just around one order of magnitude.

A feasible tuning range for today's OPAs covers a large spectrum from infrared to the ultraviolet, see e.g. \cite{OPAs}, which can in principle be further extended to 
larger frequencies by higher harmonic
generation.
Thus, asserting a tuning range of $\lambda\lesssim2\times 10^{-5} \mathrm{m}$ ($\omega \gtrsim 0.06 \mathrm{eV}$),  one could in principle explore axion mass ranges 
above $m_{\bot}\gtrsim 2.46 \mathrm{eV}$ for 
$\omega_{\mathrm{out}}=\omega_{\mathrm{in}}+\omega_{\parallel}$ 
as indicated
by the black solid vertical line in Fig. \ref{fig:bounds} for POLARIS intensities and frequency
\footnote{Here, we neglect the $m_{\parallel}$ solution, since for this mass the outgoing frequency is 
$\omega_{\mathrm{out}}=\omega_{\mathrm{in}}-\omega_{\parallel}$ and thus the frequency of the incoming probe photon cannot be reduced below $\omega_{\parallel}$.}.
We find that, employing OPAs, the exclusion bounds on ALPs could be extended to $g\gtrsim 1.8\times 10^{-4} \mathrm{GeV}^{-1}$.

However, it has to be taken care that the outgoing probe photons are shifted to frequencies outside the spectral widths of the employed lasers as discussed above.
Also, the efficiency for the OPAs and the higher harmonic generation processes have in principle to be taken into account.
In this order of magnitude estimate these effects are neglected.

In order to estimate the discovery potential at future facilities, let us extend our considerations to the planned Exawatt facility ELI \cite{ELI}.
At this facility, a potentially feasible intensity aim is at $I=10^{26} \frac{\mathrm{W}}{\mathrm{cm}^2}$.
According to the previous estimates for the available intensity, we thus obtain the ELI field strengths $E_\parallel \simeq 8.29\times 10^9 \mathrm{eV}^2$ and
$E_\bot\simeq 5.25\times 10^9 \mathrm{eV}^2$.
These field strength exceed those of the POLARIS/JETI setup by roughly two orders 
of magnitude. The achievable focal parameters can be expected to be of the same order of magnitude as in our previous estimate.

In Fig. \ref{fig:bounds} the red dotted vertical line indicates the region which could be probed at ELI with an additional OPA with tuning range and
$N_{\mathrm{in}} N_{\mathrm{shot}}\simeq10^{15}$ 
as above. Here we find that the coupling region above $g\gtrsim 1.9 \times 10^{-6} \mathrm{GeV}^{-1}$ could be explored.
Intriguingly, already a moderate demand on the number 
of interacting laser photons can almost complement the bounds of ALPS in the higher ALP mass region.
Ultimately, the red dot-dashed vertical line suggests the necessary requirements at ELI for tests of typical QCD axion models, 
which are plotted as a yellow band, see \cite{CAST} and references 
therein. In order to make contact with the range of QCD axion models, in the attainable mass region,
$N_{\mathrm{in}} N_{\mathrm{shot}}\approx 10^{26}$ is needed at ELI to explore an ALP mass range up to 
$g\gtrsim 3.4 \times 10^{-9} \mathrm{GeV}^{-1}$. This is a rather strong requirement for current technology, but with the 
advance of OPA technology and in high-intensity laser technology this could be a worthwhile long-term aim for the future.

\FIGURE
{
\centerline{
\includegraphics[scale=0.8]{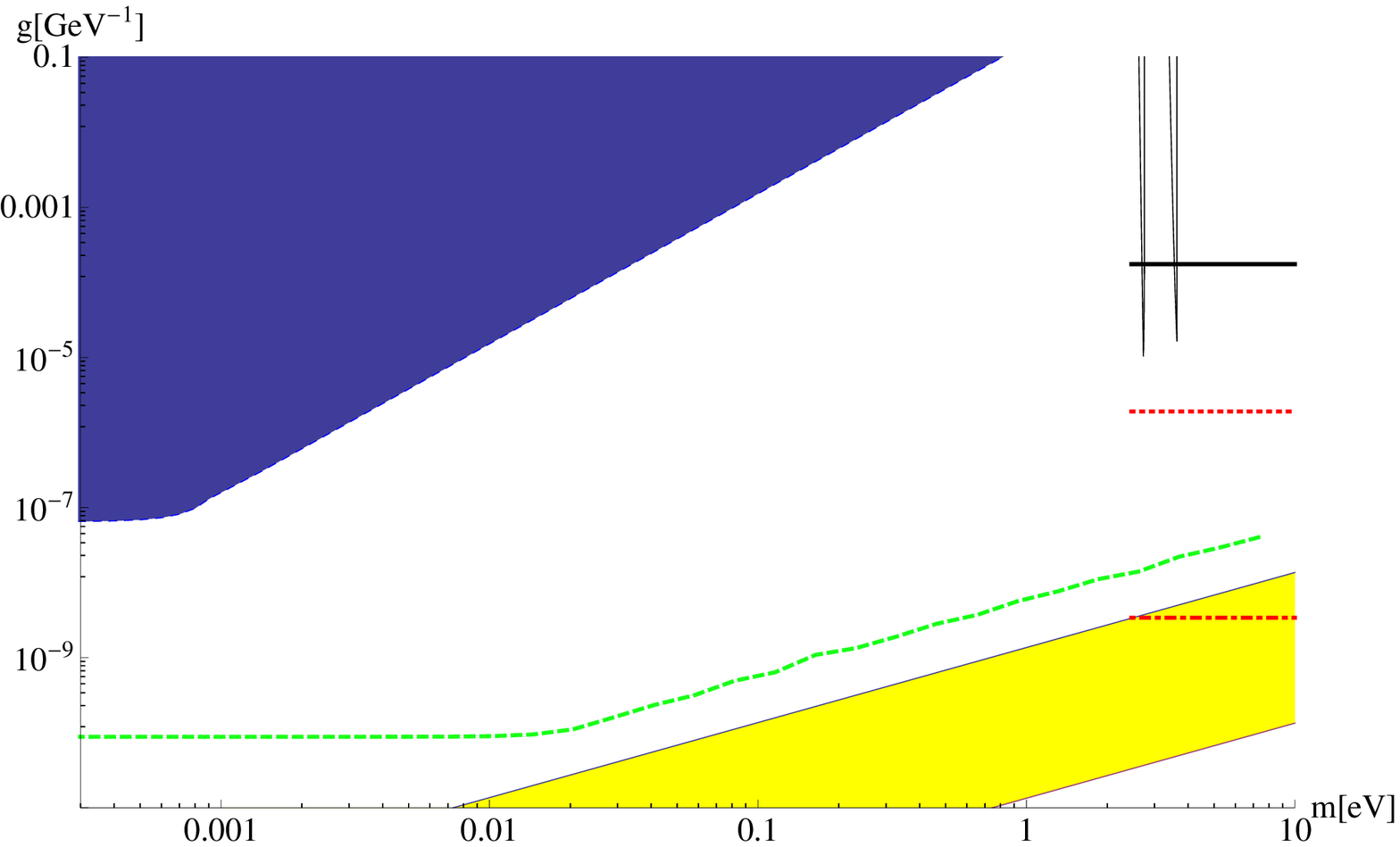}
}

\caption{Axion-like-particle exclusion bounds for purely laser-based setups in
  comparison to searches with dipole magnets, see text for parameters and
  attainable coupling values.  The blue-shaded area in the upper left corner
  gives the currently best laboratory bounds on axion-like-particles from the
  ALPS collaboration \cite{Ehret:2010mh}, while the best limits on solar
  axions are provided by the CAST experiment \cite{CAST}, denoted by a
  green-dashed line.  The black wedges denote the exclusion limits for a setup
  involving the JETI++ and POLARIS laser systems at Jena for one shot at
  single photon detection.  The black line indicates the principle exclusion
  bounds at this setup by frequency-tuning through optical parametric
  amplification. The red dotted line corresponds to an estimate for the best
  obtainable bounds with ELI with a present-day OPA system. The red dot-dashed
  line suggests the necessary requirements at ELI for testing the parameter
  regime of typical QCD axion models, which are plotted as a yellow band.}

\label{fig:bounds}
}

\section{Conclusion}
In conclusion, we have investigated the feasibility of a search for
axion-like-particles in a purely laser-based setup.  In particular, we have
concentrated on modern high-intensity laser systems, since the available field
strengths can serve as a lever arm for probing the weak coupling $g$ of
axions and axion-like-particles (ALPs) to electromagnetism.

We have studied in a one-dimensional setting the conversion of a probe beam with frequency $\omega_{\mathrm{in}}$ into an 
ALP beam in an external laser and its reconversion into a photon beam by a second external laser for Gaussian beam profiles in the formal limit of infinite pulse lengths.
The important difference to conventional photon-axion mixing in homogeneous fields arises from a split of the incoming photon into several partial waves whose frequencies are given by 
all non-negative sums and differences of the three laser frequency scales.

We have also shown that the amplitude of these partial waves is proportional to two damping terms induced by the conservation of three-momentum at the conversion points. 
The constraints imposed by momentum conservation, together with the practical requirement that the outgoing photon should be frequency shifted with respect to $\omega_{\mathrm{in}}$ for
reasons of detectability, demand for a carefully designed experimental setup.

We have shown that these constraints can be satisfied in a specific set-up involving one external beam of frequency $\omega_{\parallel}$ to counter-propagate with respect to
the probe beam and another external beam of 
frequency $\omega_{\bot}$ to propagate orthogonally to the probe beam. For the situation where  $\omega_{\bot}= 2\omega_{\parallel}$, momentum is conserved
at both conversion points at the same time while the frequency of the outgoing beam is different from the frequency of the incoming 
beam: $\omega_{\mathrm{out}}=\omega_{\mathrm{in}}+\omega_{\parallel}$ or $\omega_{\mathrm{out}}=\omega_{\mathrm{in}}-\omega_{\parallel}$ depending on the order of the 
interaction of the beams. 

This mechanism is reminiscent to sum-frequency generation and difference-frequency generation known from Nonlinear optics.

The amplitudes of theses processes are peaked around certain resonant masses which are determined by momentum conservation: 
For the frequency up-converted probe photons with $\omega_{\mathrm{out}}=\omega_{\mathrm{in}}+\omega_{\parallel}$, the resonant mass is
$m_\bot = 2 \sqrt{\omega_{\parallel}^2+\omega_{\mathrm{in}}\omega_{\parallel}}$, whereas  $m_\parallel = 2 \sqrt{\omega_{\mathrm{in}}\omega_{\parallel}}$
for the down-conversion process with $\omega_{\mathrm{out}}=\omega_{\mathrm{in}}-\omega_{\parallel}$.

We concluded that by frequency tuning by means of optical parametric amplification, purely laser-based experiments hold the prospect of providing the strongest laboratory bounds on 
axion-like-particles in the $\mathcal{O}(\mathrm{eV})$ mass range. As laboratory
searches with dipole magnets generically probe only lower mass ranges, purely
laser-based experiments could complement them in an essential manner.

In summary, our work suggests that high-intensity lasers are about to evolve into a new tool for fundamental physics.

\acknowledgments { We are grateful to M.~Nicolai, G.G.~Paulus, J.~Redondo and
  J.~Jaeckel for helpful discussions.  This work is supported by the DFG
  through grants SFB/TR18, GRK1523, and Gi 328/5-1.  }

\appendix

\section{Appendix}

In the previous calculation, we have considered the interaction of two external lasers with frequencies $\omega_\bot$ and $\omega_\parallel$ with a probe beam $\omega_{\mathrm{in}}$, 
in which the first, orthogonally propagating beam mediates the photon-axion conversion and the second, counter-propagating beam mediates the back-conversion from axions into photons. 

However, if all beams are focused simultaneously onto the same spot as considered, it is of course not experimentally obvious, which of the beams causes the conversion 
and back-conversion process, 
respectively. For this reason, we want to consider the process at interchanged interaction order $\bot \leftrightarrow \parallel$.

To this end, we employ the beam parameterization of the counter-propagating external beam, cf.  Eqs. \eqref{eq:E_k_field} and \eqref{eq:B_k_field} for the 
photon-ALP-conversion process. Now, the electric as well as the magnetic component of the external field can interact yielding an extra factor of $2$.
Then, the ALP amplitude of Eq. \eqref{eq:phi_int_int} reads
\begin{multline}
\phi(z',t')=g E_{\mathrm{in}} E_\parallel \int_{-\infty}^{\infty} \mathrm{d}z'' \frac{1}{\sqrt{1+\left(z''/z_r^{\mathrm{in}}\right)^2}} \frac{1}{\sqrt{1+\left(z''/z_r^\parallel\right)^2}} \\
\times \int_{-\infty}^{\infty}  
\mathrm{d}t'' J_{0}\left(m\sqrt{(t'-t'')^2-|z'-z''|^2}\right) \theta\left((t'-t'')-|z'-z''|\right) \\
\times
\sin\left(\omega_{\mathrm{in}} t''- k_{\mathrm{in}}z''+\arctan\left(\frac{z''}{z_r^{\mathrm{in}}}\right)\right)
\sin\left(\omega_\parallel t'' + k_\parallel z'' -\arctan\left(\frac{z''}{z_r^\parallel}\right) + \psi_\parallel\right)\ . \label{eq:phi_int_int_app}
\end{multline}
We proceed analogously to the previous calculation and perform the temporal integration by the substitution $t''\rightarrow t'-T$ and by virtue of of Eq. \eqref{eq:J0_int}. 
Specializing to $|\omega_{\mathrm{in}} \pm \omega_\parallel|>m$ and using the identity in Eq. \eqref{eq:arctan}  one finds that the equivalent to Eq. \eqref{eq:general_int} reads
\begin{multline}
\phi(z',t')=\frac{1}{4}g E_{\mathrm{in}} E_\parallel
\Biggl[\frac{i}{k_{\mathrm{ax}}^{+}}e^{i(\omega_{\mathrm{in}}+\omega_\parallel)t'}e^{-i\mathrm{sgn}(z'-z'')k_{\mathrm{ax}}^{+}z'} e^{i\psi_\parallel} \\
 \times \int_{-\infty}^{\infty} \mathrm{d}z'' \frac{1}{1- i\left(z''/z_r^{\mathrm{in}}\right)} \frac{1}{1+i\left(z''/z_r^\parallel\right)} 
e^{i(- k_{\mathrm{in}} +k_\parallel +\mathrm{sgn}(z'-z'')k_{\mathrm{ax}}^{+})z''} \\
-\frac{i\mathrm{sgn}(\omega_{\mathrm{in}}-\omega_\parallel)}{k_{\mathrm{ax}}^{-}}e^{i(\omega_{\mathrm{in}}-\omega_\parallel)t'}
e^{-i\mathrm{sgn}(z'-z'')\mathrm{sgn}(\omega_{\mathrm{in}}-\omega_\parallel)k_{\mathrm{ax}}^{-}z'} e^{-i\psi_\parallel}\\
 \times \int_{-\infty}^{\infty} \mathrm{d}z''  \frac{1}{1- i\left(z''/z_r^{\mathrm{in}}\right)} \frac{1}{1-i\left(z''/z_r^\parallel\right)} 
e^{i(- k_{\mathrm{in}}-k_\parallel+\mathrm{sgn}(z'-z'')\mathrm{sgn}(\omega_{\mathrm{in}}-\omega_\parallel)k_{\mathrm{ax}}^{-})z''} \\
+c.c.\Biggr] \ , \label{eq:general_int_app}
\end{multline}
where here, according to the substitution  $\bot \leftrightarrow \parallel$, the axion wave vector becomes a function of the frequency of the counter-propagating external field:
$k_{\mathrm{ax}}^{\pm}=\sqrt{(\omega_{\mathrm{in}}\pm \omega_\parallel)^2-m^2}$.

Again, we find the characteristic structure of the ALP partial waves with frequencies $\omega_{\mathrm{ax}}=\omega_{\mathrm{in}}\pm\omega_\parallel$, 
which have transmitted ($\mathrm{sgn}(z'-z'')=+1$) and reflected ($\mathrm{sgn}(z'-z'')=-1$) contributions.
The two remaining integrations over $z''$ can be performed in the complex $z''$-plane. The respective integrals read
\begin{eqnarray}
\int_{-\infty}^{\infty} \mathrm{d}z \frac{1}{1- i\frac{z}{z_r^{\mathrm{in}}}} \frac{1}{1+i\frac{z}{z_r^\parallel}} e^{i\delta k_\parallel^{+} z}
&=& \frac{\pi z_r^{\mathrm{in}}z_r^\parallel}{z_r^{\mathrm{in}}+z_r^\parallel} \times \nonumber \\
&{}& \Bigg[(1-\mathrm{sgn}(\delta k_\parallel^{+})) e^{\delta k_\parallel^{+}z_r^{\mathrm{in}}} \nonumber \\ 
&{}& + (1+\mathrm{sgn}(\delta k_\parallel^{+})) 
e^{-\delta k_\parallel^{+}z_r^\parallel}\Bigg]\label{eq:SFG_z_int} \\
\int_{-\infty}^{\infty} \mathrm{d}z \frac{1}{1- i\frac{z}{z_r^{\mathrm{in}}}} \frac{1}{1-i\frac{z}{z_r^\parallel}} e^{i\delta k_\parallel^{-} z}
&=& - \frac{\pi z_r^{\mathrm{in}}z_r^\parallel}{z_r^{\mathrm{in}}-z_r^\parallel} \times \nonumber \\ 
&{}& \left[(1-\mathrm{sgn}(\delta k_\parallel^{-}))\left(e^{\delta k_\parallel^{-}z_r^{\mathrm{in}}}-e^{\delta k_\parallel^{-}z_r^\parallel}\right)\right]\ , \label{eq:DFG_z_int}
\end{eqnarray}
where we have defined
\begin{eqnarray}
\delta k_\parallel^{+}&=& - k_{\mathrm{in}} +k_\parallel +\mathrm{sgn}(z'-z'')k_{\mathrm{ax}}^{+} \label{eq:delta_j_pl_app}\\
\delta k_\parallel^{-}&=& - k_{\mathrm{in}}-k_\parallel+\mathrm{sgn}(z'-z'')\mathrm{sgn}(\omega_{\mathrm{in}}-\omega_\parallel)k_{\mathrm{ax}}^{-} \ . \label{eq:delta_j_min_app}
\end{eqnarray}
In Eqs. \eqref{eq:SFG_z_int} and \eqref{eq:DFG_z_int}, we encounter again the
resonant structure of the conversion amplitude which can be
attributed to the requirement of momentum conservation.  As before, the sharp
momentum cutoffs as induced by the signum functions are in fact relaxed by an
integration over a finite interaction region.  In analogy to the previous
considerations, we determine the resonant mass $m$.

In vacuum ($\omega=k$), Eqs. \eqref{eq:delta_j_pl_app} and \eqref{eq:delta_j_min_app} become 
\begin{eqnarray}
\delta k_\parallel^{+}&=& -
\omega_{\mathrm{in}} +\omega_\parallel +\mathrm{sgn}(z'-z'')\sqrt{(\omega_{\mathrm{in}}+\omega_\parallel)^2-m^2}\stackrel{!}{\simeq}0 \label{eq:delta_j_pl_vac_app}\\
\delta k_\parallel^{-}&=& - \omega_{\mathrm{in}}-\omega_\parallel+\mathrm{sgn}(z'-z'')
\mathrm{sgn}(\omega_{\mathrm{in}}-\omega_\parallel)\sqrt{(\omega_{\mathrm{in}}-\omega_\parallel)^2-m^2}\stackrel{!}{\simeq}0 \ . \label{eq:delta_j_min_vac_app}
\end{eqnarray}
The condition in Eq. \eqref{eq:delta_j_pl_vac_app} is solved by setting $m=m_{\parallel}=2\sqrt{\omega_{\mathrm{in}} \omega_\parallel}$ in the case of 
transmission for $\omega_{\mathrm{in}} >\omega_\parallel$ and reflection for $\omega_{\mathrm{in}}<\omega_\parallel$. This can be understood intuitively: For momenta 
of the probe beam which are larger then the momenta of the counter-propagating beam, we find transmission, otherwise reflection of the ALP beam.

As in the previously considered setup, for $\delta k_\parallel^{-}$ there exists no resonant mass.
In the following, we thus again only keep the transmitted
part of the sum-frequency solution $\omega_{\mathrm{ax}}^{+}$.
A justification for omitting the reflected part will be given below Eq. \eqref{eq:delta_k_min_vac_app}.

We have as a pendant to Eq. \eqref{eq:phi_T_plus}:
\begin{multline}
\phi^{(\mathrm{T})}(z',t')\approx - \frac{1}{2} \frac{\pi z_r^{\mathrm{in}}z_r^\parallel}{z_r^{\mathrm{in}}+z_r^\parallel}g E_{\mathrm{in}} E_\parallel
[\frac{1}{k_{\mathrm{ax}}^{+}}\sin((\omega_{\mathrm{in}}+\omega_\parallel)t'-k_{\mathrm{ax}}^{+}z'+\psi_\parallel))\\
\left[(1-\mathrm{sgn}(\delta k_\parallel^{+})) e^{\delta k_\parallel^{+}z_r^{\mathrm{in}}}+(1+\mathrm{sgn}(\delta k_\parallel^{+})) e^{-\delta k_\parallel^{+}z_r^\parallel}\right] \ . \
\label{eq:phi_T_plus_app}
\end{multline}
We now turn to the back-conversion of the ALPs into photons. However, as the external beam for the back-conversion propagates orthogonally to the $z$-axis, only the 
magnetic or electric field component can couple, cf. Eq. \eqref{eq:eout}. However, due to the asymmetric coupling structure, $e_{\mathrm{out}}$ is not invariant under this choice, 
as discussed in Sec. \ref{sec:EOM}. 

As the ALP is massive, the contribution from the magnetic field component of the external beam will be larger, since it couples to the temporal derivative of $\phi$. We thus choose
$E_k^{x}=0$ 
in the following.

Following the steps below Eq. \eqref{eq:eout}, where $B_k^{y}$ is now given through \Eqref{eq:E_j_field},  we find that after the integrations over $t'$ and
$z'$ for the back-conversion, we end up with
\begin{multline}
e_{\mathrm{out}}(z,t) = -\frac{1}{16} g^2  \pi^{3/2}  \frac{z_r^{\mathrm{in}}z_r^\parallel}{z_r^{\mathrm{in}}+z_r^\parallel}  w_0^\bot E_{\mathrm{in}} E_\parallel E_\bot \frac{\omega_{\mathrm{in}}+\omega_\parallel}{k_{\mathrm{ax}}^{+}} \\
\times
\left[(1-\mathrm{sgn}(\delta k_\parallel^{+})) e^{\delta k_\parallel^{+}z_r^{\mathrm{in}}}+(1+\mathrm{sgn}(\delta k_\parallel^{+})) e^{-\delta k_\parallel^{+}z_r^\parallel}\right] \\
\times
\Biggl[\frac{1}{i} e^{i (\omega_{\mathrm{in}}+\omega_\parallel+\omega_\bot)(t-\mathrm{sgn}(z-z')z)}e^{i(\psi_\parallel+\psi_\bot)} e^{-\frac{1}{4} (w_0^\bot \delta k_\bot^{+})^2} \\-
\frac{1}{i} e^{i (\omega_{\mathrm{in}}+\omega_\parallel-\omega_\bot)(t-\mathrm{sgn}(z-z')z)}e^{i(\psi_\parallel-\psi_\bot)} e^{-\frac{1}{4} (w_0^\bot \delta k_\bot^{-})^2}  +c.c.\Biggr]\ , 
\label{eq:e_out_int_app}
\end{multline}
with the abbreviations
\begin{eqnarray}
\delta k_\bot^{+}&=& -k_{\mathrm{ax}}^{+} + \mathrm{sgn}(z-z'') (\omega_{\mathrm{in}}+\omega_\parallel+\omega_\bot)  \label{eq:delta_k_pl_app}\\
\delta k_\bot^{-}&=& -k_{\mathrm{ax}}^{+} + \mathrm{sgn}(z-z'') (\omega_{\mathrm{in}}+\omega_\parallel-\omega_\bot) \ . \label{eq:delta_k_min_app}
\end{eqnarray}
In Eq. \eqref{eq:e_out_int_app}, we encounter the familiar behavior of the reconverted photons. The outgoing electromagnetic wave is composed of two partial waves whose amplitude is 
tied to the vanishing of the sum of the momenta in the process: $\delta k_\bot^{\pm}$. As before, we are interested in the situation of concurrent momentum conservation in both
conversion processes, under the constraint $\omega_{\mathrm{out}}\neq\omega_{\mathrm{in}}$. Thus, we rewrite 
Eqs. \eqref{eq:delta_k_pl_app} and \eqref{eq:delta_k_min_app} as
\begin{eqnarray}
-\delta k_\bot^{+}&=& \sqrt{(\omega_{\mathrm{in}}+\omega_\parallel)^2-m^2} - \mathrm{sgn}(z-z') (\omega_{\mathrm{in}}+\omega_\parallel+\omega_\bot) \stackrel{!}{\simeq} 0  
\label{eq:delta_k_pl_vac_app}\\
-\delta k_\bot^{-}&=& \sqrt{(\omega_{\mathrm{in}}+\omega_\parallel)^2-m^2} - \mathrm{sgn}(z-z') (\omega_{\mathrm{in}}+\omega_\parallel-\omega_\bot) \stackrel{!}{\simeq} 0  \ , 
\label{eq:delta_k_min_vac_app}
\end{eqnarray}
where we have multiplied the equations by $-1$, which is justified since $\delta k_\bot^{\pm}$ appears as a square in Eq. \eqref{eq:e_out_int_app}. 

By comparing the above conditions for momentum conservation in the back-conversion process to the condition for the photon-ALP conversion in Eq. \eqref{eq:delta_j_pl_vac_app}, 
we again find, that for positive frequencies only $\delta k_\bot^{-}$ can 
be ''matched'' to the vanishing of $\delta k_\parallel^{+}$ for transmission and the relation $\omega_\bot= 2\omega_\parallel$.
As the frequency of the outgoing beam evaluates to $\omega_{\mathrm{out}}=\omega_{\mathrm{in}}-\omega_\parallel$ in this situation,
we again find that only the transmitted parts of both waves can propagate undamped, since the solution requires $\omega_{\mathrm{in}}>\omega_\parallel$
(cf. discussion below Eq. \eqref{eq:delta_j_min_vac_app}).

In summary, we find for the outgoing electromagnetic wave at $\omega_\bot= 2\omega_\parallel$ the transmitted part
\begin{multline}
e_{\mathrm{out}}^{(\mathrm{T})}(z,t) \approx \frac{1}{8} g^2  \pi^{3/2}  \frac{z_r^{\mathrm{in}}z_r^\parallel}{z_r^{\mathrm{in}}+z_r^\parallel} 
w_0^\bot E_{\mathrm{in}} E_\parallel E_\bot \frac{\omega_{\mathrm{in}}+\omega_\parallel}{k_{\mathrm{ax}}^{+}}
e^{-\frac{1}{4} \left(w_0^\bot \delta k\right)^2} \\
\left[(1-\mathrm{sgn}(\delta k)) e^{\delta k z_r^{\mathrm{in}}}+(1+\mathrm{sgn}(\delta k )) e^{-\delta k z_r^\parallel}\right]  
\sin((\omega_{\mathrm{in}}-\omega_\parallel)(t-z)+\psi_\parallel-\psi_\bot) \ , \label{eq:e_out_par_bot_app}
\end{multline}
where we have set $\delta k_\parallel^{+}=\delta k_\bot^{-}=\delta k$.

\end{document}